\documentclass[
superscriptaddress,
amsmath,amssymb,
aps,
prl,
twocolumn,
]{revtex4-2}

\usepackage{graphicx}
\usepackage[dvipsnames]{xcolor}
\usepackage{bm}
\usepackage{siunitx}
\usepackage{booktabs}
\usepackage[colorlinks=true, linkcolor=blue, citecolor=blue, urlcolor=blue]{hyperref}
\usepackage{notes2bib}

\newcommand{\Td}{T_{\mathrm{d}}}

\newcommand{\bk}{\mathbf{k}}

\begin{document}

\title{Fourth-order Optoelectronic Response from Cascaded Circular Photogalvanic and Nonlinear Hall Effects}

\author{Bhupendra Sharma}
\affiliation{Department of Mechanical Engineering, University of Rochester,
Rochester, New York 14627, USA}

\author{Sobhit Singh}
\email{s.singh@rochester.edu}
\affiliation{Department of Mechanical Engineering, University of Rochester,
Rochester, New York 14627, USA}
\affiliation{Materials Science Program, University of Rochester, Rochester, New York 14627, USA}

\begin{abstract}
The interplay between nonlinear optical transitions and topological band structure offers a route to control photocurrents. 
We reveal a fourth-order optoelectronic response that emerges due to an interlink between the circular photogalvanic effect (CPGE) and the Berry curvature dipole (BCD) in noncentrosymmetric 2D materials. 
Using monolayer $\Td$-WTe$_2$ as a prototype, we predict that circularly polarized mid-infrared light produces a steady dc injection current that induces an internal electric field, which in turn drives a transverse nonlinear Hall response through BCD. The resulting cascaded photovoltage scales as the fourth power of the optical field $E_0^4$.
By mapping the full injection current tensor, we show that this cascaded voltage is strongly tunable by the optical geometry: normal incidence drives an in-plane resonance $\mathrm{Im}(\eta_{yxy})$, whereas oblique illumination ($\theta = 45^{\circ}$) recruits a dominant out-of-plane component $\mathrm{Im}(\eta_{yyz})$ and amplifies the signal by more than two orders of magnitude (${\sim}10^2~\mu$V). 
While the massive linear Drude background typically screens nonlinear responses in semimetals, we argue that the amplitude modulation of the optical pump allows lock-in detection to cleanly isolate the frequency-doubled cascaded response. 
The proposed mechanism converts mid-infrared light into a gate-tunable transverse signal, providing a route for probing quantum geometry and realizing topological photodetectors and frequency doublers.
\end{abstract}


\maketitle

The nonlinear response of quantum materials to electromagnetic fields has become a central tool for probing the geometry of electronic Bloch states~\cite{jiang2025revealing,orenstein2021topology}. The quantum geometric tensor, whose imaginary part gives the Berry curvature and real part yields the quantum metric~\cite{provost1980riemannian,sala2026probing}, encodes the information governing these responses at successive orders in the applied field. At first order, the Berry curvature produces the anomalous Hall effect~\cite{nagaosa2010anomalous}. At second order, the Berry curvature dipole (BCD) induces a nonlinear Hall effect (NLHE)~\cite{sodemann2015quantum,ma2019observation,xu2018electrically, Singh_PRL2020}, while the circular photogalvanic effect (CPGE) generates helicity-dependent injection photocurrents that are linked to the interband Berry connection~\cite{de2017quantized,sipe2000second, rees2020helicity}. Third-order responses driven by Berry curvature and quantum metric quadrupoles have been further observed in several materials~\cite{yu2025quantum,li2024quantum,mandal2024quantum,liu2025giant,fang2024quantum}.

\begin{figure}[t]
\centering
\includegraphics[scale=0.2]{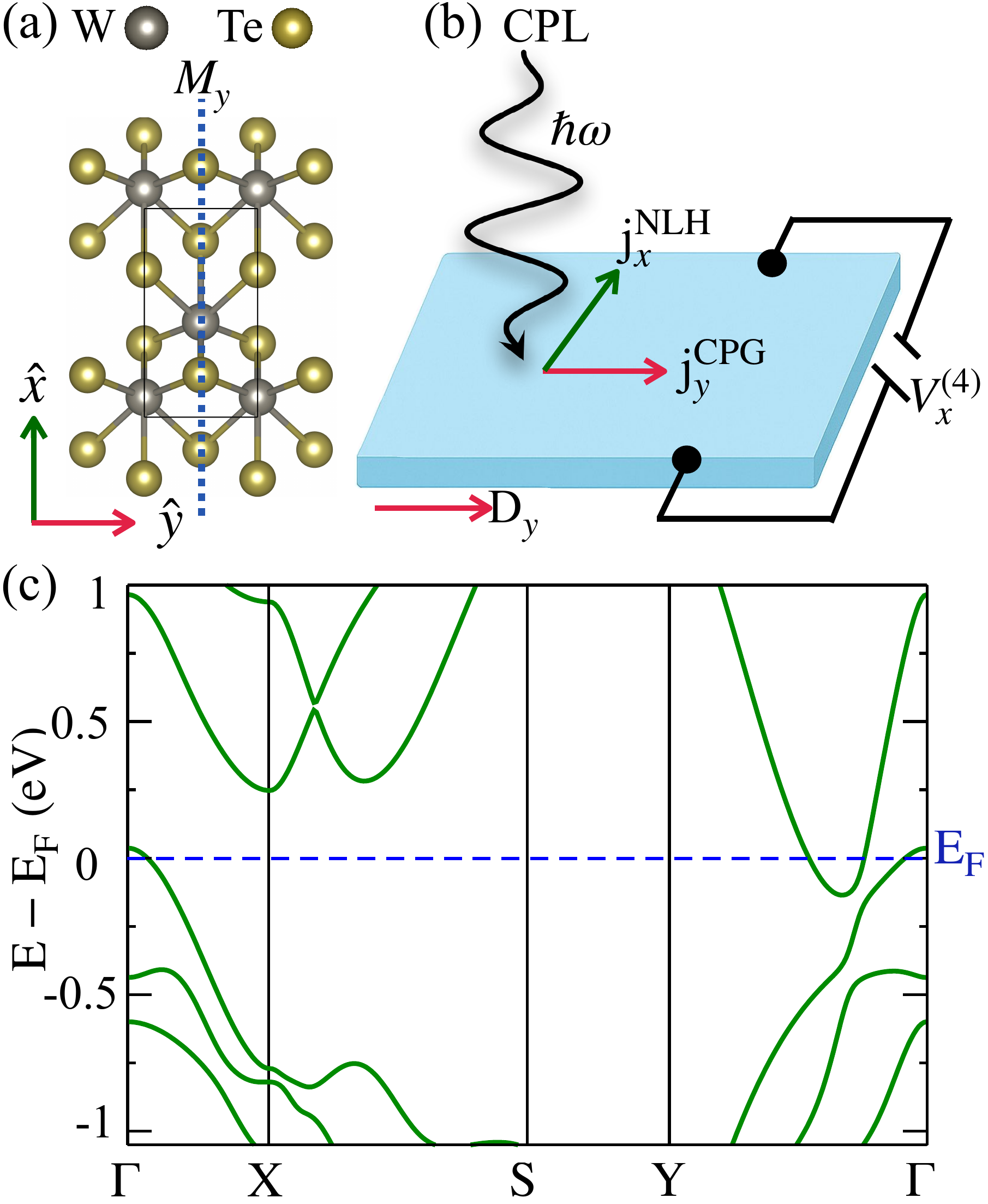}
\caption{(a) Crystal structure of monolayer $\Td$-WTe$_2$. 
(b) Schematic illustration of the fourth-order mechanism: circularly polarized light (CPL) injects a CPGE photocurrent $j_y^{\mathrm{CPG}}$ along $\hat{y}$; the open-circuit internal field $E_y^{\mathrm{int}}$ drives a transverse nonlinear Hall current $j_x^{\mathrm{NLH}}$ along $\hat{x}$ via the ${D}_y$; which induces fourth-order voltage drop $V_x^{(4)}$ across $\hat{x}$. (c)  Electronic band structure of $\Td$-WTe$_2$ near the Fermi level ($E_F$).}
\label{BCD+CPGE}
\end{figure}

A natural question is whether fourth-order responses can be realized and measured. Direct four-photon processes require exceedingly high field strengths~\cite{boyd2008nonlinear, shen1983principles, monoszlai2020measurement}. 
A cascaded approach, i.e., combining two second-order processes sequentially, can instead achieve an effective fourth-order response at moderate field amplitudes. 
We recognize that both the BCD-driven NLHE and the CPGE share a common origin rooted in the Berry curvature, and can coexist in the same class of noncentrosymmetric, time-reversal-invariant (TRI) materials. 
The CPGE-generated photocurrent creates an internal built-in dc electric field that acts as the input for the subsequent NLHE, producing a transverse voltage proportional to fourth power of the optical field $E_0^4$. We develop a microscopic theory of this cascaded CPGE$\rightarrow$NLHE response and apply it to monolayer $\Td$-WTe$_2$.

The proposed CPGE$\rightarrow$NLHE cascade establishes a fundamentally distinct route to nonlinear Hall physics. 
While conventional BCD-induced NLHE measurements rely on low-frequency electrical excitation (typically 10-1000\,Hz \cite{du2021nonlinear,qin2024light}) 
\bibnote{The BCD induced NLHE is measured using low-frequency electrical currents (typically 10–1000\,Hz) because higher frequencies rapidly diminish the signal.},
our mechanism harnesses mid-infrared optical fields (20-30\,THz) to generate a nonlinear Hall voltage through an entirely internal optoelectronic process. 
Furthermore, because the polarity of the response directly reflects the sign of the BCD, the final readout can be continuously tuned via electrostatic gating of the chemical potential. 
Moreover, the strong dependence of the signal on photon energy and illumination geometry provides a direct access to the underlying quantum geometry of Bloch bands. These features transform the NLHE from a transport probe into an optical spectroscopic tool, opening opportunities for quantum-geometric imaging, topological photodetection, and high-frequency nonlinear optoelectronic functionalities.



\textit{Symmetry and the CPGE injection tensor.}--- Monolayer $\Td$-WTe$_2$ crystallizes in the monoclinic space group Pm (No.~6). The point-group generators are the mirror $\mathcal{M}_y$, time reversal $\mathcal{T}$, and their product $\mathcal{M}_y\mathcal{T}$; inversion symmetry is broken, which is the prerequisite for any second- or fourth-order response. 
Under $\mathcal{M}_y$ the coordinates transform as $(x,y,z)\to(x,-y,z)$, so a rank-$n$ polar tensor component with $N_y$ indices equal to $y$ acquires a factor $(-1)^{N_y}$ 
\bibnote{Under $\mathcal{M}_y$ mirror, every $y$-index contributes a minus sign, while $x$ and $z$ indices do not. Hence, a rank-$n$ polar tensor component with $N_y$ indices equal to $y$ transforms with an overall factor of  $(-1)^{N_y}$},
and invariance requires $N_y$ to be even. 
The CPGE injection current is the second-order dc response to circularly-polarized light (CPL) whose monochromatic field satisfies $\mathbf{E}(-\omega)=\mathbf{E}^*(\omega)$~\cite{de2017quantized}. For CPL the antisymmetric part $\mathbf{E}(\omega)\times\mathbf{E}^*(\omega)\neq 0$, and the photocurrent rate $dj_a/dt$ is governed by the imaginary part of the rank-3 injection conductivity tensor $\eta_{abc}$~\cite{sipe2000second}, with $\{a,b,c\}$ spanning the Cartesian coordinates $\{x,y,z\}$. Because $\mathrm{Im}(\eta_{abc})$ is antisymmetric in its last two indices under time-reversal symmetry (TRS), it can be contracted with the Levi-Civita symbol to define a rank-2 pseudotensor
\begin{equation}
  \beta_{ab} = \epsilon_{bkl}\,\mathrm{Im}(\eta_{akl}),
  \label{eq:beta_def}
\end{equation}
so that the CPGE injection current rate takes the form~\cite{de2017quantized}
\begin{equation}
  \frac{dj_a}{dt} = \beta_{ab}\,\bigl[\mathbf{E}(\omega)\times\mathbf{E}^*(\omega)\bigr]_b.
  \label{eq:CPGE_pseudo}
\end{equation}
Then, $\beta_{ab}$ is determined by the quantum geometry of Bloch bands ~\cite{de2017quantized},
\begin{equation}
  \beta_{ab}(\omega)
  = \frac{\pi e^3}{\hbar V}\,\epsilon_{bkl}
    \sum_{\mathbf{k},n,m}
    f_{nm}^{\mathbf{k}}\,
    \Delta_{\mathbf{k},nm}^{a}\,
    r_{\mathbf{k},nm}^{k}\,
    r_{\mathbf{k},mn}^{l}\,
    \delta(\hbar\omega - \varepsilon_{\mathbf{k},mn}),
  \label{eq:beta_micro}
\end{equation}
where $V$ is the cell volume, $f_{nm}^{\mathbf{k}}$ is the Fermi--Dirac occupation difference, $r_{\mathbf{k},nm}^{k}=i\langle n|\partial_{k}|m\rangle$ denotes the interband Berry connection, and $\Delta_{\mathbf{k},nm}^{a}$ is the group-velocity difference between bands $n,m$ with energy gap $\varepsilon_{\mathbf{k},mn}$ at crystal momentum $\mathbf{k}$. 
Imposing TRS and $\mathcal{M}_y$, four independent antisymmetric pairs of $\mathrm{Im}(\eta_{abc})$ are allowed, which are listed in Table~\ref{tab:injection}.

\textit{Incidence geometry.}---For CPL propagating along $\hat{z}$ (i.e., the normal incidence) the helicity vector is $\mathbf{h} = \eta\, E_0^2\,\hat{z}$, where $\eta = \pm 1$ is the photon helicity~\cite{ganichev2003spin}. Only $h_z$ contributes, and the in-plane photocurrent is $\left.dj_y/dt\right|_{\theta=0} = \beta_{yz}\,\eta\, E_0^2$. No current flows along $\hat{x}$ because $\beta_{xz}\propto \mathrm{Im}(\eta_{xxy})$ is $\mathcal{M}_y$-forbidden. For an oblique beam tilted by $\theta$ from $\hat{z}$ within the $xz$-plane~\cite{bai2025circular}, the helicity vector acquires an in-plane component $h_x = \eta E_0^2 \sin\theta$, activating $\beta_{yx}$:
\begin{equation}
\frac{dj_y}{dt}
= [\beta_{yz}\cos\theta + \beta_{yx}\sin\theta]\,\eta\, E_0^2,
\label{eq:oblique}
\end{equation}
with $\beta_{yx} \propto \mathrm{Im}(\eta_{yyz})$. The total CPGE current along $\hat{y}$ thus combines two contributions weighted by the incidence angle and we are going to exploit this angle dependence to optimize the cascade response.

\textit{BCD-induced NLH conductivity.}---Within the semiclassical Boltzmann framework, the BCD reads~\cite{sodemann2015quantum}
\begin{equation}
D_{bd} = \int \frac{d^2k}{(2\pi)^2}\; f_0(\bk)\;
\frac{\partial \Omega_d(\bk)}{\partial k_b},
\end{equation}
where, $f_0$ is the equilibrium distribution and $\Omega_d(\bk)$ is the Berry curvature. The induced nonlinear Hall conductivity is~\cite{sodemann2015quantum,zhang2018berry}
\begin{equation}
    \chi_{abb} = -\epsilon_{adb} \frac{e^3 \tau D_{bd}}{2\hbar^2 (1+i\omega \tau)},
    \label{nlh_conductivity}
\end{equation}
where, $\tau$ is the scattering time and $\epsilon_{adb}$ is the Levi-Civita symbol with $a,b=x,y$ and $d=z$ in 2D~\cite{zhang2018electrically}. Under $\mathcal{M}_y$ the only allowed 2D component is $D_{y} \equiv D_{yz}$ (Fig.~\ref{BCD+CPGE}).

\begin{figure}[t]
\centering
\includegraphics[scale=0.5]{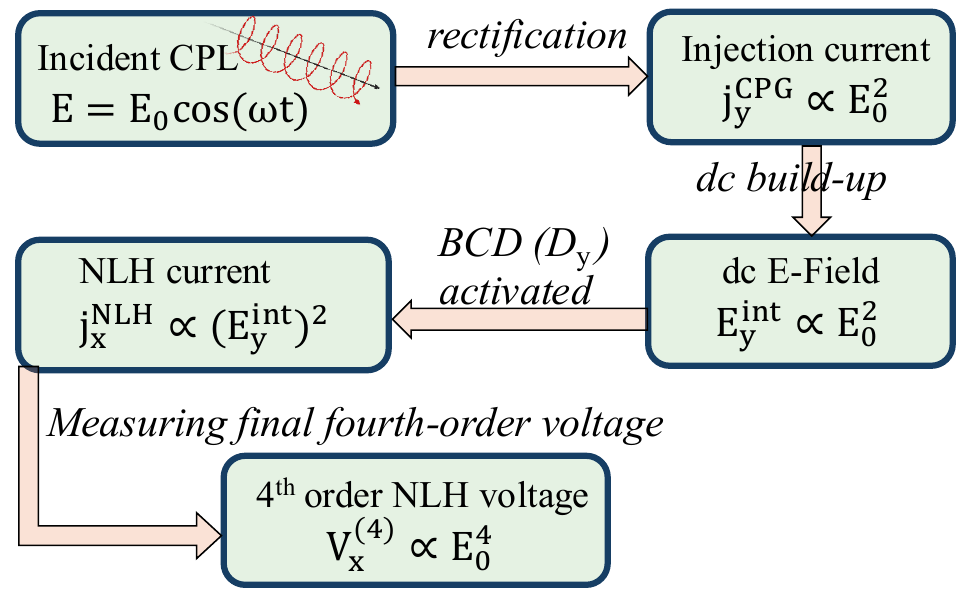}
\caption{Schematic of the fourth-order cascaded optoelectronic response. 
The $\hat{x}$ and $\hat{y}$ directions lie along the crystallographic $\mathbf{a}$ and $\mathbf{b}$ axes.}
\label{fig:cascade_schematics}
\end{figure}

\textit{Cascaded fourth-order response.}---We now derive the fourth-order voltage that emerges from the sequential coupling of the CPGE injection current and the BCD-driven NLHE as schematically illustrated in Fig.~\ref{fig:cascade_schematics}. 
The saturated injection photocurrent along $\hat{y}$~\cite{de2017quantized} is: $j_y^{\mathrm{CPG}} = \beta_{\mathrm{eff}}\,\eta\, E_0^2$, with the effective CPGE coefficient
\begin{equation}
\beta_{\mathrm{eff}}(\omega,\theta)
= \beta_{yz}\cos\theta + \beta_{yx}\sin\theta
\label{eq:beta_eff}
\end{equation}
In an open-circuit steady state, the photocurrent is balanced by an opposing Ohmic current, $j_y^{\mathrm{Ohm}} + j_y^{\mathrm{CPG}} = 0$, 
producing the internal field~\cite{belinicher1980photogalvanic,sturman2021photovoltaic}
\begin{equation}
E_y^{\mathrm{int}} = -\frac{\beta_{\mathrm{eff}}\,\eta\, E_0^2}{\sigma_{yy}},
\label{eq:Eint}
\end{equation}
where $\sigma_{yy}$ is the longitudinal Drude conductivity. 
The internal field further drives a transverse nonlinear Hall current via the BCD component $D_y$~\cite{sodemann2015quantum},
\begin{equation}
j_x^{\mathrm{NLH}} = \chi_{xyy}\,(E_y^{\mathrm{int}})^2,\quad
\chi_{xyy} = \frac{e^3\,\tau\, D_{y}}{2\hbar^2\,(1 + i\omega_m\tau)},
\label{eq:jNLH}
\end{equation}
with the amplitude modulation frequency $\omega_m$, to be distinguished from the incident optical frequency $\omega$. Combining Eqs.~(\ref{eq:Eint})--(\ref{eq:jNLH}) and dividing by $\sigma_{xx}$ for the open-circuit voltage across length $l_x$ yields our main result,
\begin{equation}
\boxed{V_x^{(4)} =
\frac{e^3\,\tau\, D_{y}\,\beta_{\mathrm{eff}}^2\, l_x}
{2\,\hbar^2\,\sigma_{yy}^2\,\sigma_{xx}\,(1+i\omega_m\tau)}\;
E_0^4.}
\label{eq:VoltageFinal}
\end{equation}
The voltage scales as $E_0^4$, realizing a fourth-order response obtained through a cascade of two second-order processes (Fig.~\ref{fig:cascade_schematics}). The sign of $V_x^{(4)}$ follows $D_{y}$, enabling possible topological switching via Fermi-level tuning. 

In the dc limit ($\omega_m\tau \ll 1$),
the voltage response is linearly dependent on $\tau$, whereas for $\omega_m\tau \gg 1$ the response becomes $\tau$-independent and rolls off as $|V_x^{(4)}|\propto 1/\omega_m$; see Supporting Materials (SM)~\cite{SM} for details.

\begin{figure}[t]
	\centering
	\includegraphics[scale=0.15]{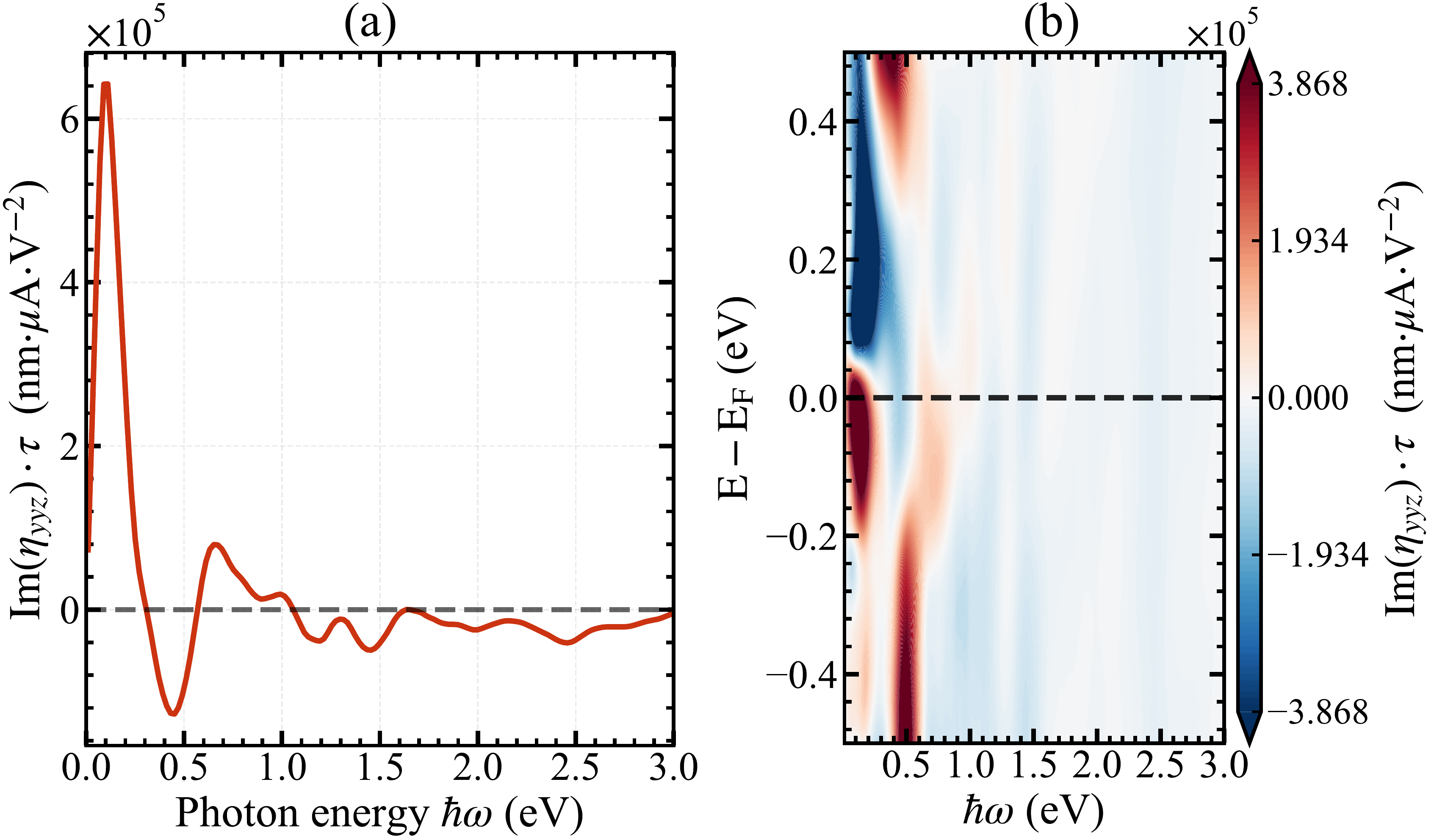}
	\caption{(a) CPGE injection-current spectrum $\mathrm{Im}(\eta_{yyz})$ for $\tau = 5$~ps. This component drives the built-in electric field at oblique incidence; the resonance at $\hbar\omega \approx 0.11$~eV reaches ${\sim}6.4\times 10^{5}$~nm$\cdot\mu$A/V$^{2}$. (b) Colormap of $\mathrm{Im}(\eta_{yyz})$ in nm$\cdot\mu$A/V$^{2}$ at $\tau = 5$~ps. The dashed line marks the Fermi energy ($E_F$).}
    \label{fig:yyz_spectra_heatmap}
\end{figure}

\textit{CPGE Injection current.}---Table~\ref{tab:injection} summarizes the symmetry-allowed components of the CPGE injection current tensor  $\mathrm{Im}(\eta_{abc})$ at the Fermi level. All components with an odd number of $y$-indices are numerically zero, confirming the $\mathcal{M}_y$ selection rule, and the TRS antisymmetry $\mathrm{Im}(\eta_{abc}) = -\mathrm{Im}(\eta_{acb})$ holds to machine precision. The normal-incidence component $\mathrm{Im}(\eta_{yxy})$ peaks at $3.8\times 10^{4}$~nm$\cdot\mu$A/V$^{2}$ at $\hbar\omega = 0.09$~eV, agreeing with the quantum-kinetic value $7.14\times 10^{4}$~nm$\cdot\mu$A/V$^{2}$ of Liu \textit{et al.}~\cite{liu2024photogalvanic} under gating ($\nu = 30$~GHz, $\varepsilon_F = 0.08$~eV, $\tau = 5$~ps, $E_\perp = 1$~V/nm). The oblique component $\mathrm{Im}(\eta_{yyz})$ peaks an order of magnitude higher, $6.4\times 10^{5}$~nm$\cdot\mu$A/V$^{2}$ at $\hbar\omega = 0.11$~eV. The separate resonances indicate distinct interband transitions, providing independent optical channels to pump the intermediate static field $E_y^{\text{int}}$. Fig.~\ref{fig:yyz_spectra_heatmap}(a) shows the frequency dependence of $\mathrm{Im}(\eta_{yyz})$ at the Fermi level, and 
Fig.~\ref{fig:yyz_spectra_heatmap}(b) demonstrates the gate-tunability of the optical pump stage (see SM~\cite{SM} for $\mathrm{Im}(\eta_{yxy})$ part).

\begin{table}[t]
\caption{Symmetry-allowed components of the injection current tensor $\mathrm{Im}(\eta_{abc})$ in monolayer $\Td$-WTe$_2$, reported as 2D sheet conductivities at $\tau = 5$~ps, with the activated CPGE pseudotensor component $\beta_{ab}$.}
\label{tab:injection}
\begin{ruledtabular}
\begin{tabular}{lccc}
TRS pair &  $|\mathrm{Im}(\eta_{abc})|_{\mathrm{peak}}$ &
$\hbar\omega_{\mathrm{peak}}$ & $\beta_{ab}$ \\
 & (nm$\cdot\mu$A/V$^{2}$) & (eV) & \\
\colrule
$[\mathrm{Im}(\eta_{yxy}),\mathrm{Im}(\eta_{yyx})]$  & $3.8\times 10^{4}$ & 0.09 & $\beta_{yz}$ \\
$[\mathrm{Im}(\eta_{yyz}),\mathrm{Im}(\eta_{yzy})]$  & $6.4\times 10^{5}$ & 0.11 & $\beta_{yx}$ \\
$[\mathrm{Im}(\eta_{xxz}),\mathrm{Im}(\eta_{xzx})]$ &  $2.8\times 10^{4}$ & 0.45 & $\beta_{xy}$ \\
$[\mathrm{Im}(\eta_{zxz}),\mathrm{Im}(\eta_{zzx})]$  & $2.3\times 10^{5}$ & 0.55 & $\beta_{zy}$ \\
\end{tabular}
\end{ruledtabular}
\end{table}

\begin{figure}[t]
	\centering
   \includegraphics[scale=0.15]{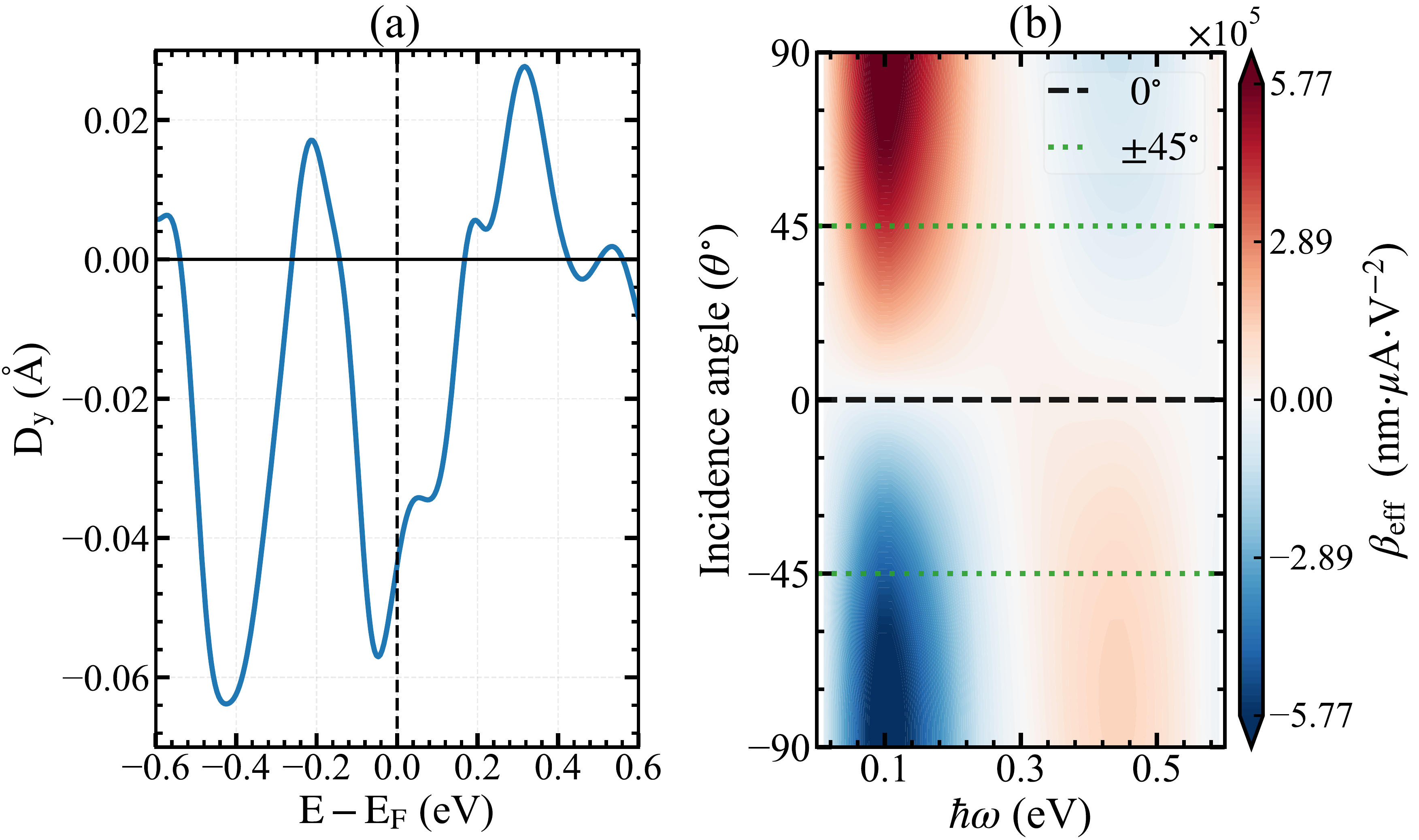}
	\caption{(a) The BCD component $D_{y}$ of monolayer $\Td$-WTe$_2$ as a function of chemical potential near $E_F$. 
    (b) Colormap of the effective CPGE generation rate $\beta_{\mathrm{eff}}(\theta, \omega)$ versus incidence angle $\theta$ and photon energy $\hbar\omega$. Horizontal lines indicate standard optical geometries ($0^\circ$, $\pm 45^\circ$). 
    Traversing this $(\theta, \hbar\omega)$ phase space continuously tunes and maximizes the cascade efficiency by angular or spectral switching.}
    \label{fig:bcd_angMap}
\end{figure}

\textit{Transducing the cascaded voltage.}---While the injection current $j_y^{\rm{CPG}}$ and longitudinal Drude conductivity $\sigma_{yy}$ set the strength of the built-in electric field $E_y^{\text{int}}$, the BCD component $D_y$ transduces the static $E_y^{\text{int}}$ into the final 
dc voltage $V_x^{(4)}$. The two cascade steps operate at very different frequencies, and this separation is what makes the scheme work. The CPGE rectifies the optical field ($\hbar\omega \approx 0.09$--$0.11$~eV, i.e.\ $\omega/2\pi \approx 22$--$27$~THz) into a dc current that sets up $E_y^{\rm int}$. 
From Eq.~(\ref{nlh_conductivity}), the BCD conductivity $\chi^{\rm BCD}(\omega)\propto i/(\omega+i\Gamma)$ with $\Gamma=1/\tau$ is largest at zero frequency and damps as $1/\omega$ once $\omega\tau\gtrsim 1$. 
Liu \textit{et al.}~\cite{liu2024photogalvanic} reported that the BCD term dominates the nonlinear Hall response of $\Td$-WTe$_2$ up to ${\sim}1$~THz, so the dc induced $E_y^{\rm int}$ lets the BCD operate within its dominant window. The $(1+i\omega_m\tau)^{-1}$ part in Eq. (\ref{eq:VoltageFinal}) is related to the amplitude modulation frequency of the laser $\omega_m$, which rolls off near $\omega_m\tau\sim 1$ ($\nu_m\approx 32$~GHz for $\tau=5$~ps). Fig.~\ref{fig:bcd_angMap}(a) shows ${D}_y$ as a function of chemical potential near the Fermi level calculated on a $1000\times 1000\times 1$ k-grid: ${D}_{y} \approx -0.04$~\AA\ at $E_F$ is consistent with prior reports~\cite{ma2019observation, you2018berry, zhang2018electrically}, peaking at $|{D}_{y}| \approx 0.07$~\AA\ for $E-E_F\approx-0.4$~eV (see SM~\cite{SM} for BCD convergence details). Since $V_x^{(4)} \propto {D}_{y}$, the sign reversal of ${D}_{y}$ across $E_F$ makes the output polarity electrically switchable.

\textit{Incidence-angle dependence and spectral switching.}---The final transverse voltage scales with the square of the intermediate photovoltage, and thus with the square of the effective CPGE conductivity: $V_x^{(4)} \propto (E_y^{\mathrm{int}})^2 \propto \beta_{\mathrm{eff}}^2$. The components $\mathrm{Im}(\eta_{yxy})$ and $\mathrm{Im}(\eta_{yyz})$ shows resonance peaks at different photon energies (0.09~eV and 0.11~eV), so the incidence angle acts as a \textit{spectral selector} through the trigonometric weighting of the generation rate:
\begin{equation}
\beta_{\mathrm{eff}}(\theta,\omega) = \mathrm{Im}(\eta_{yxy})(\omega)\cos\theta
+ \mathrm{Im}(\eta_{yyz})(\omega)\sin\theta.
\label{eq:beta_eff_spec}
\end{equation}

At normal incidence ($\theta = 0^{\circ}$), only the $\mathrm{Im}(\eta_{yxy})$ term is activated, yielding a resonant peak at $\hbar\omega = 0.09$ eV. Consequently, the voltage scales as $V_x^{(4)} \propto \mathrm{Im}(\eta_{yxy})^2$. Once the optical incident direction tilts away from normal, the out-of-plane $\mathrm{Im}(\eta_{yyz})$ contribution switches on. Bai \textit{et al.}~\cite{bai2025circular} show that macroscopic CPGE photocurrents in topological semimetals can be optimized by sweeping the incidence angle between $-45^{\circ}$ and $45^{\circ}$ within the $xz$-plane. The resonant peak of $\mathrm{Im}(\eta_{yyz})$ ($\hbar\omega = 0.11$ eV, 26.6 THz) is preferred to maximize the CPGE conductivity because it exceeds the $\mathrm{Im}(\eta_{yxy})$ component by nearly an order of magnitude. For instance, tilting the incident 26.6 THz laser to an angle of $\theta = 45^{\circ}$ enhances the final fourth-order voltage according to,
\begin{multline}
V_{\text{oblique}}^{(4)}(45^{\circ}) \propto \left[ \mathrm{Im}(\eta_{yxy})\cos 45^{\circ} + \mathrm{Im}(\eta_{yyz})\sin 45^{\circ} \right]^2 \\
= \tfrac{1}{2}\left(\mathrm{Im}(\eta_{yxy}) + \mathrm{Im}(\eta_{yyz})\right)^2,
\label{eq:Voblique}
\end{multline}
and comparing with normal incidence at the same frequency gives the geometric enhancement
\begin{equation}
\frac{V_{\text{oblique}}(45^{\circ})}{V_{\text{normal}}(0^{\circ})}
= \frac{\tfrac{1}{2}\left(\mathrm{Im}(\eta_{yxy}) + \mathrm{Im}(\eta_{yyz})\right)^2}{\mathrm{Im}(\eta_{yxy})^2}
\approx 127.
\label{eq:ratio}
\end{equation}
As $\theta \to 90^{\circ}$, $\sin\theta \to 1$ and the $\mathrm{Im}(\eta_{yxy})$ contribution vanishes, so the cascaded voltage approaches its maximum, $V_{\max}^{(4)} \propto \mathrm{Im}(\eta_{yyz})^2$. However, experimentally approaching $90^{\circ}$ is unfeasible. At grazing incidence, optical excitation is hampered by near-total Fresnel reflection and the spatial elongation of the laser beam footprint, both of which decimate the effective optical power density reaching the sample \cite{hecht2001eugene}. Hence, an incidence angle of roughly $45^{\circ}$ remains the optimal ``sweet spot'' that maximizes the output photocurrent in 2D architectures \cite{boland2023narrowband, ni2020linear, cheng2024chirality}. Because the two channels connect continuously across the $(\hbar\omega,\theta)$ space [Fig.~\ref{fig:bcd_angMap}(b)], the cascade can be tuned by \textit{angular switching} (i.e., fix $\omega$, tilt sample between the $\mathrm{Im}(\eta_{yxy})$ and $\mathrm{Im}(\eta_{yyz})$ resonances) or \textit{spectral switching} (i.e., fix $\theta \sim 45^{\circ}$, sweep $\omega$).

\textit{Fourth-order voltage estimates.}---At normal incidence and the $\mathrm{Im}(\eta_{yxy})$ resonance ($\hbar\omega = 0.09$~eV), $V_x^{(4)} \approx 0.94~\mu$V, evaluated with parameters for a 2D system:
$\sigma_{xx} \approx \sigma_{yy} \approx 10^{-4}$~S, $l_x = 1~\mu$m, ${D}_{y} = 0.07$~\AA, $\tau = 5$~ps, and $E_0 = 10^{5}$~V/m. Conventional DC Hall devices typically operate on the millivolt scale, burying sub-microvolt responses~\cite{crescentini2021hall}. However, by amplitude-modulating the optical pump and employing an AC lock-in amplifier (noise floor ${\sim}2.5\text{ nV}/\sqrt{\text{Hz}}$~\cite{zhinst_hall}), the signal is lifted clear of the background and resolved using standard setups~\cite{kang2019nonlinear}. The oblique incidence at the $\mathrm{Im}(\eta_{yyz})$ resonance ($\hbar\omega = 0.11$~eV, $\theta = 45^{\circ}$) produces the geometric enhancement as discussed above. Under identical parameters, $V_x^{(4)} \approx 119~\mu$V in the dc limit, rising to ${\sim}187~\mu$V at $\theta = 60^{\circ}$. An output at the $2\omega_m$ harmonic arises from the NLH step which squares the amplitude modulated intermediate field~\cite{ma2017direct, kang2019nonlinear, ma2019observation}. 

One might expect the signal to be severely suppressed at high frequencies since the voltage decays as $V_x^{(4)} \propto 1/\omega_m$. However, even at frequency $\nu_m = 100\text{ GHz}$ ($\omega_m\tau \approx 3.1$), the output persists at $\approx 36\ \mu\text{V}$ at a $45^{\circ}$ incidence~\cite{SM}. 
This shows that the proposed mechanism can function as an ultrafast topological frequency doubler~\cite{chen2020anisotropic}. Additionally, as $V_x^{(4)} \propto {D}_{y}$, materials with larger BCD linearly scale the output voltage response; monolayer $\Td$-WMoTe$_4$ with $|{D}_{y}| \approx 4.5$~\AA~\cite{99gy-krxw} (${\sim}64\times$ that of $\Td$-WTe$_2$), projects $V_x^{(4)}$ to ${\sim}7.6$~mV at $\theta = 45^{\circ}$.

\textit{Background-free, gate-tunable mid-infrared photodetection.}---The cascaded response separates the symmetry-allowed topological signals from the polarization-independent Drude and photo-thermoelectric backgrounds that limit infrared detection using semimetals~\cite{osterhoudt2019colossal}. 
Due to the sign reversal of the $D_y$ across $E_F$ [Fig.~\ref{fig:bcd_angMap}(a)], electrostatic gating provides a straightforward way to reverse the output polarity without changing the device or optical geometry~\cite{xu2018electrically}, a capability missing in conventional photodiodes~\cite{liu2021silicon}. Furthermore, the injection resonances span the mid- to near-infrared ($0.09$--$0.55$~eV, $2$--$14~\mu$m) range, overlapping the molecular-fingerprint window, which fits high-resolution chemical imaging~\cite{bhargava2012infrared}, selective biological~\cite{baker2014using} and trace-gas~\cite{hodgkinson2013optical,tittel2003mid} sensing, and thermal imaging~\cite{rogalski2012history,corsi2010history}, and quantum sensing~\cite{isobe2020high}.

In summary, we predict a fourth-order optoelectronic response arising from an intrinsic coupling between the CPGE and the BCD-induced NLHE in time-reversal-invariant, noncentrosymmetric materials.
The cascade mechanism exploits a CPGE-generated dc injection current to drive nonlinear Hall transport, thereby circumventing the frequency limitations of conventional nonlinear Hall measurements and enabling topological transport to be pumped by mid-infrared light.
The resulting transverse voltage scales as $E_0^4$ and directly reflects the magnitude and sign of the BCD, allowing continuous control through electrostatic gating.
Our first-principles calculations on monolayer $\Td$-WTe$_2$ map the injection current tensor and reveal its strong sensitivity to the optical measurement geometry. 
Our results uncover an intriguing interplay between topological optical excitation and nonlinear quantum transport, establishing 
a platform for probing quantum geometry, topological photodetection, and ultrafast nonlinear optoelectronics.


\begin{acknowledgments}
The authors acknowledge support from the Furth Research Fund at the University of Rochester, and from the U.S.~Department of Energy, Office of Science, Office of Fusion Energy Sciences, Quantum Information Science Program, under Award No. DE-SC0020340.
\end{acknowledgments}

\bibliography{references}

\end{document}